\documentclass[10pt,letterpaper,twocolumn,prl,showpacs]{revtex4-1}

\usepackage[english]{babel}

\usepackage{amsmath,amssymb}
\usepackage[dvips]{graphicx,color}

\begin{document}

\title{Writing and erasing of temporal cavity solitons\\ by direct phase modulation of the cavity driving field}

\author{Jae K. Jang,$^*$ Miro Erkintalo, Stuart G. Murdoch, and St\'{e}phane Coen}

\affiliation{Dodd-Walls Centre for Photonic and Quantum Technologies, and Physics Department, The University of
Auckland, Private~Bag~92019, Auckland 1142, New Zealand\\ $^*$Corresponding author: jake.jang.ur.mate@gmail.com}

\begin{abstract}%
  \noindent Temporal cavity solitons (CSs) are persisting pulses of light that can manifest themselves in
  continuously driven passive resonators, such as macroscopic fiber ring cavities and monolithic microresonators.
  Experiments so far have demonstrated two techniques for their excitation, yet both possess drawbacks in the form
  of system complexity or lack of control over soliton positioning. Here we experimentally demonstrate a new CS
  writing scheme that alleviates these deficiencies. Specifically, we show that temporal CSs can be excited at
  arbitrary positions through direct phase modulation of the cavity driving field, and that this technique also
  allows existing CSs to be selectively erased. Our results constitute the first experimental demonstration of
  temporal cavity soliton excitation via direct phase modulation, as well as their selective erasure (by any
  means). These advances reduce the complexity of CS excitation and could lead to controlled pulse generation in
  monolithic microresonators.
\end{abstract}

\maketitle

\noindent Cavity solitons (CSs) are solitary waves that can persist in driven nonlinear passive
resonators~\cite{ackemann_chapter_2009}. Their tendency to broaden is balanced by nonlinear self-focusing, and all
energy they lose is replenished  by the coherent field driving the cavity. Several of them can simultaneously
co-exist, and they can be independently excited, erased and manipulated~\cite{firth_optical_1996,
brambilla_interaction_1996, barland_cavity_2002, jang_temporal_2014}. These properties have led to CSs being
identified as promising candidates for bits in all-optical buffers and processing units~\cite{mcdonald_spatial_1990,
wabnitz_suppression_1993, leo_temporal_2010}.

Historically the focus has been on spatial CSs, i.e., non-diffracting localized beams of light trapped in planar
cavities~\cite{brambilla_interaction_1996, barland_cavity_2002, firth_optical_1996}. More recently experiments
performed in  optical fiber loops~\cite{leo_temporal_2010, jang_ultraweak_2013, jang_observation_2014,
jang_temporal_2014} have stimulated interest in their temporal counterparts: recirculating
non-dispersive pulses of light~\cite{wabnitz_suppression_1993}. Such temporal CSs do not suffer from material defects
that hinder the performance of spatial cavities~\cite{caboche_microresonator_2009}, although somewhat analogous
impairments can arise due to electrostriction-mediated interactions~\cite{jang_ultraweak_2013}. These interactions
can however be overcome by phase modulating the continuous wave (cw) laser driving the cavity; the CSs are then
trapped to the peaks of the ensuing intracavity phase profile~\cite{firth_optical_1996, jang_temporal_2014}. In
parallel with investigations in \emph{macroscopic} fiber cavities, temporal CSs have also attracted attention in the
context of monolithic \emph{microresonators}~\cite{herr_temporal_2014,  saha_modelocking_2013}. In these devices, CSs
can under certain conditions correspond to the temporal structures underlying broadband ``Kerr'' frequency
combs~\cite{coen_modeling_2013, erkintalo_coherence_2014}, suggesting applications in metrology and high repetition
rate pulse train generation.

Excitation of CSs requires that the resonator cw steady-state is suitably perturbed~\cite{mcdonald_spatial_1990,
mcdonald_switching_1993, hachair_cavity_2005, barbay_incoherent_2006}. For temporal CSs two techniques have been
demonstrated. Experiments in fiber resonators have used an incoherent ``writing'' scheme~\cite{leo_temporal_2010,
jang_ultraweak_2013, jang_observation_2014, jang_temporal_2014}, where an optical ``addressing'' pulse perturbs the
cw steady-state via nonlinear cross-phase modulation (XPM). In microresonators, stable soliton states have been
reached by carefully tuning the frequency of the cavity driving field~\cite{herr_temporal_2014,
saha_modelocking_2013}. Both methods yield CSs, but also possess drawbacks: optical addressing increases the system
complexity~\cite{leo_temporal_2010}, whilst adjusting the driving laser frequency provides limited control over how
many CSs are  excited and at what temporal positions~\cite{erkintalo_coherence_2014}. Moreover, neither of these
techniques can easily be adapted to achieve controlled erasure of existing CSs. XPM has been numerically proposed for
this purpose~\cite{leo_temporal_2010}, but difficulties in synchronization have prevented experimental realization.
This represents a major shortcoming, since the potential to be independently erased is widely regarded as a defining
characteristic of CSs, underpinning many of their proposed functionalities~\cite{barland_cavity_2002,
hachair_cavity_2005, barbay_incoherent_2006}.

Here, we implement another method for temporal CS excitation that alleviates these deficiencies. Our technique relies
on the phase modulation that is typically applied to the cavity driving field to trap CSs into specific time-slots
\cite{jang_temporal_2014}. By applying local boosts in the phase modulation, we are able to write CSs into
corresponding empty slots. Significantly, similar boosts also allow us to demonstrate selective erasure of CSs
already trapped in the cavity. This technique minimizes the complexity of CS systems, since the same components
enable writing, trapping, and erasure. It could also represent a step towards controlled pulse generation in optical
microresonators~\cite{taheri_soliton_2014}.

\begin{figure}[t]
  \includegraphics[width=1\columnwidth, clip = true]{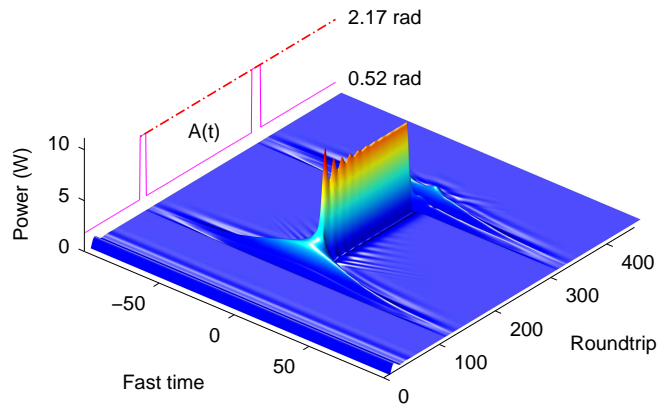}
  \caption{Numerical results of CS writing and erasing by means of abrupt changes in the phase of the cavity driving
    field. The phase modulation amplitude $A(t)$ is shown as the curve on the left. The numerical parameters are
    similar to the experiments that follow: $P_\mathrm{in} = 908~\mathrm{mW}$, $t_\mathrm{R} = 0.48~\mu\mathrm{s}$, $L = 100$~m,
    $\theta = 0.1$, $\alpha = 0.146$, $\gamma = 1.2~\mathrm{W^{-1}km^{-1}}$, $\beta_2 = -21.4~\mathrm{ps^2/km}$,
    $\delta_0 = 0.4426$~rad, and $\tau_0 = 43$~ps.}
  \label{simulation}
\end{figure}

The ability to switch dispersive cw bistable devices through appropriate changes of the phase of the driving field
was first proposed by Hopf \emph{et al.}~\cite{hopf_anomalous_1979}, and analyzed theoretically in the context of CSs
by McDonald and Firth~\cite{mcdonald_switching_1993}. Phase modulation has also been used recently to excite
localized topological phase structures in a laser with external forcing and feedback \cite{garbin_topological_2015}.
For a better understanding of our experiment, we first illustrate the phase-induced switching dynamics of temporal
CSs by means of numerical simulations. To this end, we model the evolution of the field envelope $E(t,\tau)$ inside a
high-finesse fiber resonator using the well-known mean-field equation~\cite{wabnitz_suppression_1993,
leo_temporal_2010, coen_modeling_2013}:
\begin{multline}\label{LLequation}
t_R \frac{\partial E}{\partial t} = \left(-\alpha + i \gamma L |E|^2 - i \delta_0
- i \frac{\beta_2 L}{2} \frac{\partial^2}{\partial \tau^2} \right)E(t,\tau)\\
+ \sqrt{\theta}\, S(t,\tau).
\end{multline}
Here, $t$ is a slow time variable that describes evolution over consecutive roundtrips and $\tau$ is a fast time that
describes the temporal profile of the field envelope. $t_\mathrm{R}$ is the cavity roundtrip time, $\alpha$ is half
the fraction of power lost per roundtrip, $\gamma$ is the nonlinearity coefficient,  $L$ is the cavity length and
$\beta_2$  is the group-velocity dispersion coefficient. The parameter $\delta_0$ describes the phase detuning of the
driving field $S(t,\tau)$ from the nearest cavity resonance and $\theta$ is the power transmission coefficient used
to couple light into the resonator. The driving field is cw with power $P_\mathrm{in}$, periodically phase-modulated
with a one-bit pattern synchronized to the cavity free-spectral range, $S(t,\tau) =
P_\mathrm{in}^{1/2}\mathrm{exp}[i\phi(t,\tau)]$, where $\phi(t,\tau) = A(t)\mathrm{exp}(-\tau^2/\tau_0^2)$. Our
simulations use parameters similar to the experiments that follow (see caption of Fig.~\ref{simulation}),

Figure~\ref{simulation} illustrates how CS writing and erasing can be achieved by dynamically controlling the phase
modulation amplitude $A(t)$. Starting with a cold cavity, we turn on the driving field with a weak Gaussian [70~ps
full-width at half maximum (FWHM)] phase modulation, $A = 0.52~\mathrm{rad}$, which mimics that used in our
experiments to overcome soliton interactions~\cite{jang_ultraweak_2013, jang_temporal_2014}. As can be seen, this
weak phase modulation alone does not lead to CS excitation. Instead, after some tens of roundtrips the field
converges to a cw steady-state. After 100 roundtrips we abruptly boost the phase modulation amplitude to $A =
2.17$~rad for 10 roundtrips. This leads to a broad peak in the \emph{optical intensity} which evolves and settles
into a $2.6$~ps wide (FWHM) CS in about 100 roundtrips. The CS then persists until we again boost the phase
modulation amplitude to $2.17$~rad for 15 roundtrips, which erases the CS after a brief transient.

\begin{figure}[b]
  \includegraphics[width=1\columnwidth, clip = true]{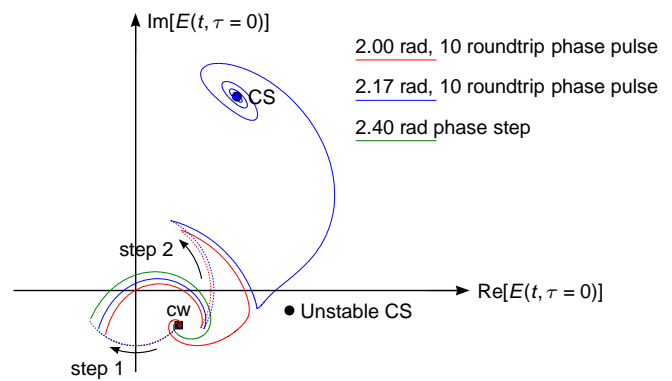}
  \caption{Simulated phase-space dynamics during attempts of CS writing with three different phase-modulation
    sequences.}
  \label{phase_space}
\end{figure}
The ability to write and erase using phase modulation can be understood by noting that an abrupt step in the driving
phase is equivalent to rotating the intracavity field in complex phase-space around the
origin~\cite{hopf_anomalous_1979}. Writing (erasing) occurs if the sequence of rotations, combined with the
intervening nonlinear evolution, is such that the field falls within the basin of attraction of a single CS
(lower-state cw). This of course implies that not all phase operations lead to switching, as is illustrated for CS
writing in Fig.~\ref{phase_space}. Here we plot the simulated evolution of the amplitude $E(t,\tau=0)$ in the complex
plane (with phase calculated relative to that of the driving field at $\tau=0$), starting from the cw solution, for
three different phase sequences. The blue curve corresponds to the successful writing procedure in
Fig.~\ref{simulation}, whereby the phase modulation amplitude is first boosted to $2.17$~rad (the corresponding
rotation is labeled as step~1) and then reduced back to $0.52$~rad after 10 roundtrips (step~2). As can be seen, the
field eventually spirals to the stable CS solution (this spiraling represents the transient oscillations seen in
Fig.~\ref{simulation}). In contrast, if the phase modulation amplitude during the 10-roundtrip boost is slightly
lower ($2.00$~rad, red curve), no writing takes place; the field remains in the basin of attraction of the original
cw solution. Similar behavior is observed when the modulation amplitude is only boosted but never returned to its
original value, and the green curve in Fig.~\ref{phase_space} shows a trajectory for this case. We note, however,
that these dynamics are particular to our choice of parameters, and for others, writing can be achieved even with a
single phase-step~\cite{mcdonald_switching_1993}. In fact, the precise dynamics depends quite sensitively on the
parameters involved. This may be linked to the metastable nature of the unstable CS solution (which exists for the
same parameters~\cite{leo_temporal_2010}), and to which the emerging soliton field is initially attracted to (see
blue curve in Fig. \ref{simulation}), as well as to the associated non-critical
slowing~\cite{mcdonald_switching_1993}.  We find that, for given parameters, some trial and error is required to
determine phase sequences suitable for writing and erasing. Fortunately this task is quite effortless; whilst the
precise dynamics may somewhat differ, switching nevertheless occurs over a wide range of conditions. We must remark
that CS switching can also be realized with abrupt changes to other parameters, even with a static phase profile.
Such changes can be induced by, e.g., mechanically perturbing the cavity.

\begin{figure}[b]
  \includegraphics[width=1\columnwidth, clip = true]{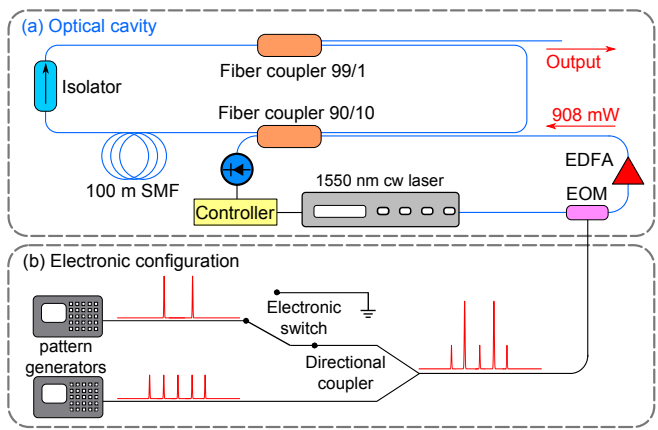}
  \caption{Schematic illustration of the (a) optical and (b) electronic segments of the experimental setup.}
  \label{setup}
  \vskip -3mm
\end{figure}
To experimentally demonstrate CS writing and erasing, we use the set-up schematically illustrated in
Fig.~\ref{setup}. The optical cavity is similar to that used in~\cite{jang_ultraweak_2013, jang_temporal_2014},
consisting of 100~meters of single-mode fiber (SMF) closed on itself with a 90/10 fiber coupler. The cavity is driven
with a narrow linewidth cw laser at 1550~nm whose output is phase-modulated with an electro-optic modulator (EOM) and
then amplified with an erbium-doped fiber amplifier (EDFA). An electronic feedback control loop is used to actively
lock the laser frequency near a cavity resonance. The cavity also hosts  a fiber isolator, used to suppress
stimulated Brillouin scattering, as well as a 99/1 tap-coupler, through which the intracavity dynamics are monitored
with a 12~GHz real-time oscilloscope. Before detection, the cavity output is filtered using a narrow (0.6~nm width)
bandpass filter (BPF) centered at 1551~nm. This improves the signal-to-noise ratio by removing the cw
component~\cite{leo_temporal_2010}.

The abrupt phase changes that enable switching are achieved by manipulating the electronic signal that drives the EOM
[see Fig.~\ref{setup}(b)]. Two programmable pattern generators, capable of producing approximately 70 ps FWHM
electronic pulses, are synchronised by a single external clock, such that the repetition rate of their output
patterns coincide with the cavity free-spectral range. One of them [top in Fig.~\ref{setup}(b)] is set to selectively
produce pulses whose amplitude is $\sim 3$~times larger than those from the other. The low-amplitude pattern is fed
to the EOM at all time; this gives rise to an intracavity phase profile that allows CSs to be
trapped~\cite{jang_temporal_2014}, but not written or erased. In contrast, the high-amplitude pattern is controlled
by an electronically gated switch. When the switch is activated, the outputs of both pattern generators are combined
for about 10~roundtrips. In this way the amplitudes of those electronic pulses that are contained in the
high-amplitude pattern can be selectively boosted to $\sim 4$~times their original level.

\begin{figure}[b]
  \includegraphics[width=1\columnwidth, clip = true]{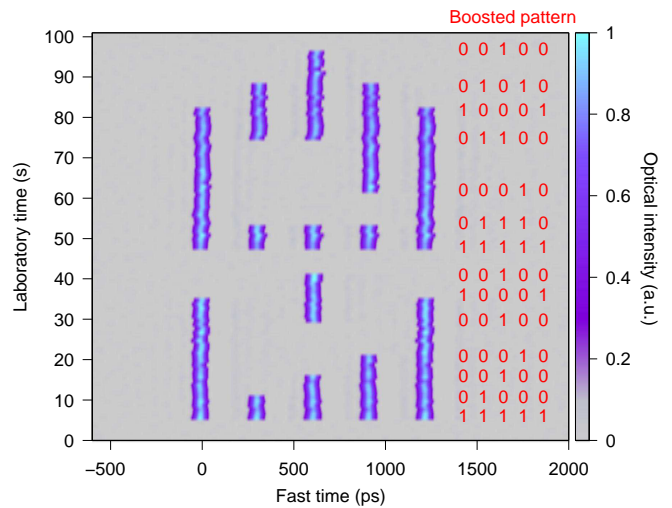}
  \vskip -2mm
  \caption{Experimental density map showing successive oscilloscope traces of the optical intensity at the cavity output.
    All the traces were acquired at 1~frame/s with a 40~GSa/s real-time oscilloscope. The bit sequences on the right
    indicate which of the five phase pulses are boosted to achieve writing and erasing. }
  \label{experiment}
  \vskip -2mm
\end{figure}
For a particular experimental demonstration, we set the low-amplitude pattern generator to produce a sequence of 5
electronic pulses separated by 300~ps (the whole pattern repeats once per cavity roundtrip). We then explore complex
writing and erasing sequences by selectively boosting some of the amplitudes as described above. Experimental results
are shown in Fig.~\ref{experiment}. The density map concatenates successive oscilloscope traces recorded at the
cavity output, illustrating how the optical intensity inside the resonator is affected by abrupt changes in phase
modulation amplitude (the bit sequences on the right indicate which electronic pulses are boosted in amplitude).
During the first 5 seconds of measurement no CSs are excited, confirming that the low-amplitude modulation itself is
insufficient for this purpose. At $t = 5~\mathrm{s}$ all five electronic pulses are boosted for 10 roundtrips,
resulting in the creation of five CSs. After excitation, the CSs remain trapped to the peaks of the low-amplitude
phase modulation~\cite{firth_optical_1996, jang_temporal_2014}. We then demonstrate erasure of individual CSs, by
boosting selected phase pulses already trapping a CS, and highlight the flexibility of the scheme by performing
complex simultaneous writing and erasing operations. In this context we remark that we are experimentally able to
write and erase using the same operation (same phase amplitude applied over the same number of roundtrips). This
should be contrasted with simulations of Fig.~\ref{simulation}, where different number of roundtrips were required
for writing and erasure. We believe this discrepancy arises due to environmental fluctuations that alter the system
parameters during the measurement. This notion is supported by the fact that we do not achieve 100\,\% fidelity, but
instead several attempts are occasionally required for successful switching.

\begin{figure}[t]
  \includegraphics[width=1\columnwidth, clip = true]{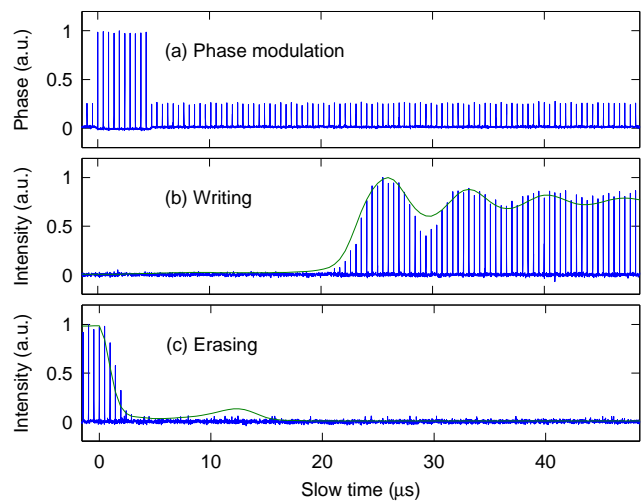}
  \vskip -2mm
  \caption{Real-time dynamics of CS writing and erasing. (a) Electronic signal fed to the EOM. (b, c) Experimentally measured roundtrip-to-roundtrip dynamics of (b) writing and (c) erasing. The green
  curves show results from numerical simulations.}
  \label{transient}
  \vskip -3mm
\end{figure}

The above experiment demonstrates that direct phase modulation allows temporal CSs to be written and erased. However,
the acquisition rate in this measurement was limited to 1~frame/s, which is too slow to capture the
switching transients. To gain more insights, we have thus also taken real-time measurements that resolve the
roundtrip-to-roundtrip dynamics of the optical intensity at the cavity output together with  the electronic signal
that drives the EOM. As above, we use the same phase operation for both writing and erasing (except that we use here
only one electronic pulse per roundtrip), and Fig.~\ref{transient}(a) shows the corresponding electronic signal.
Experimental results for transient writing and erasing dynamics are shown in Figs.~\ref{transient}(b) and (c),
respectively. During writing, the field displays ringing similar to that observed during optical
excitation~\cite{leo_temporal_2010} and numerical simulations (Fig.~\ref{simulation}), and stabilizes in about 100
roundtrips after the moment of addressing. Erasure, on the other hand, appears to take place almost immediately. For
comparison, we also plot in Figs.~\ref{transient}(b) and (c) results from numerical simulations (green curves),
taking into account the effects of the BPF and the limited bandwidth of our photodetector. The simulations use
parameters approximated from experimental measurements (see caption of Fig.~\ref{simulation}), yet for erasure we
allowed 5\,\% increase in the cavity detuning $\delta_0$ so that both writing and erasure can be achieved with the
same experimentally used phase sequence. We believe this is justified given the imperfections in the electronics used
to stabilize the laser frequency. Regardless, the numerical results show good qualitative agreement with experiments.
In this context, we also note that the writing (erasing) dynamics appear slower (faster) than they truly are since
the optical BPF does not capture the broad temporal feature that appears during switching (see
Fig.~\ref{simulation}).

To conclude, we have experimentally demonstrated writing and erasing of temporal CSs by direct manipulation of the
phase profile of the cavity driving field. The demonstrated technique offers a means to simplify the implementation
of temporal CS-based all-optical buffers, and could enable controlled pulse train generation in monolithic
microresonators. In this context, we must emphasize that writing is achieved with phase pulses that have a much
longer duration than the CSs (70~ps versus $2.6$~ps). Our work also reports the first experimental demonstration of
temporal CS erasure. This highlights that temporal CSs are truly independent entities, and constitutes a step towards
their use as fully addressable bits in optical buffers.

\end{document}